\documentclass[aps,twocolumn,amsmath,letterpaper]{revtex4}
\usepackage{amssymb}
\usepackage{graphicx}
\usepackage{array}
\usepackage{hhline}
\usepackage{longtable}

\begin{document}

    \title{Chaotic behavior of a spin-glass model on a Cayley tree}

    \author{F. A. da Costa$^{1}$, J. M. de Ara\' ujo$^{1}$ and S. R. Salinas$^{2}$}
 \affiliation{$^1$ Departamento de F\'{\i}sica Te\'orica e Experimental, Universidade Federal do Rio Grande do Norte, 59078-970, Natal, RN, Brazil}
    \affiliation{$^2$ Instituto de F\'{\i}sica, Universidade de S\~{a}o Paulo,\\05508-000, S\~{a}o Paulo, SP, Brazil}

\begin{abstract}

We investigate the phase diagram of a spin--1 Ising spin-glass model on a
Cayley tree. According to early work of Thompson and collaborators, this
problem can be formulated in terms of a set of nonlinear discrete recursion
relations along the branches of the tree. Physically relevant solutions
correspond to the attractors of these mapping equations. In the limit of
infinite coordination of the tree, and for some choices of the model
parameters, we make contact with findings for the phase diagram of more
recently investigated versions of the Blume-Emery-Griffiths spin-glass model.
In addition to the anticipated phases, we numerically characterize the existence of
modulated and chaotic structures.



\end{abstract}

    \maketitle
    \def\s{\rule{0in}{0.28in}}
    \setlength{\LTcapwidth}{\columnwidth}

In the beginning of the 1980s, Inawashiro, Frankel, and Thompson
\cite{Inawashiro81}, used a formalism of distribution functions, as developed
by a number of authors \cite{Katsura79}, to analyze the effects of disorder on
the phase diagram of a magnetic lattice gas model. This pioneering work
remained almost unknown, although the disordered magnetic lattice gas (DMLG)
is similar to some versions of the spin-$1$ Blume-Emery-Griffiths spin-glass
(BEGsg) model, which have been the subject of several investigations, both
using the replica method, at a mean-field level \cite{Crisanti02}, and using
real-space renormalization-group techniques \cite{Ongun08}.

Thompson and collaborators \cite{Thompson86} have also formulated the DMLG
model on a Cayley tree \ of coordination $z$. The physical solutions, in the
\textquotedblleft deep interior\textquotedblright\ of the tree, come from the
analysis of the attractors of a set of coupled discrete nonlinear recursion
relations. In the infinite-coordination limit, $z\rightarrow\infty$, in which
one regains the usual replica-symmetric solutions of spin-glass models, this
mapping is considerably simplified. Thompson and collaborators established the
connections between the solutions on the Cayley tree and previous calculations
on the basis of distribution functions. However, as far as we know, there has
been no attempt to carry out a more detailed analysis of the nonlinear
recursion relations, for some representative sets of model parameters, and to
make a contact with more recent findings for the BEGsg model.

We then decided to revisit the BEGsg model on the Cayley tree, in the limit of
infinite coordination, using the recursion relations obtained by Thompson and
collaborators \cite{Thompson86}. In these spin-$1$ models, the competition
between spin-glass and uniform quadrupolar terms may lead to fixed points and
limit cycles, as it has been suggested by a number of investigations for
infinite-range spin-glass models \cite{daCosta97}\cite{Katayama00}. Therefore,
we consider a particular version of the original model, which is chosen to
include this type of competition. In the infinite-coordination limit, the
discrete non-linear mapping associated with this particular version is
two-dimensional and becomes amenable to a detailed analysis. In the
low-temperature regime, and for sufficiently repulsive biquadratic spin
interactions, we find a rich behavior, including chaotic structures,
characterized by positive Lyapunov exponents.

Consider a random spin-$1$ Hamiltonian with bilinear and biquadratic terms,%
\begin{equation}
\mathcal{H}=-%
{\displaystyle\sum\limits_{\left(  i,j\right)  }}
J_{ij}S_{i}S_{j}-%
{\displaystyle\sum\limits_{\left(  i,j\right)  }}
U_{ij}S_{i}^{2}S_{j}^{2},
\end{equation}
where $S_{i}=+1,0,-1$, for all lattice sites, and the sums are over
nearest-neighbor pairs of spin variables on a Cayley tree of coordination $z$.
We assume that $\left\{  J_{ij}\right\}  $ and $\left\{  U_{ij}\right\}  $ are
independent, identically distributed quenched random variables, of Gaussian
form, with mean values $\left\langle J_{ij}\right\rangle =J_{0}/z$ and
$\left\langle U_{ij}\right\rangle =K/z$, and mean square deviations,
$\left\langle \left(  \Delta J_{ij}\right)  ^{2}\right\rangle =J^{2}/z$ and
$\left\langle \left(  \Delta U_{ij}\right)  ^{2}\right\rangle =U^{2}/z$. In
the limit of infinite coordination, $z\rightarrow\infty$, the recursion
relations have already been obtained by Thompson and collaborators (see
equations 3.30 to 3.35 of Ref. \cite{Thompson86}). We now assume a particular
case, with $J_{0}=0$ (pure spin-glass) and $U=0$ (randomness is restricted to
bilinear spin interactions). It is then easy to use the results of Ref.
\cite{Thompson86} to write a set of two-dimensional recursion relations along
the successive generations of the Cayley tree,%
\begin{equation}
q_{j+1}=%
{\displaystyle\int\limits_{-\infty}^{+\infty}}
M_{j}^{2}\left(  x\right)  \exp\left(  -\frac{1}{2}x^{2}\right)  \frac
{dx}{\sqrt{2\pi}},\label{eq3}%
\end{equation}%
\begin{equation}
p_{j+1}=%
{\displaystyle\int\limits_{-\infty}^{+\infty}}
P_{j}\left(  x\right)  \exp\left(  -\frac{1}{2}x^{2}\right)  \frac{dx}%
{\sqrt{2\pi}},\label{eq4}%
\end{equation}
where%
\begin{equation}
M_{j}\left(  x\right)  =\frac{2\sinh\left(  \beta Jq_{j}^{1/2}x\right)
}{Z_{j}\left(  x\right)  },
\end{equation}%
\begin{equation}
P_{j}\left(  x\right)  =\frac{2\cosh\left(  \beta Jq_{j}^{1/2}x\right)
}{Z_{j}\left(  x\right)  },
\end{equation}
with%
\begin{equation}
Z_{j}\left(  x\right)  =\exp\left[  -\beta Kp_{j}-\frac{1}{2}\left(  \beta
J\right)  ^{2}\left(  p_{j}-q_{j}\right)  \right]  +2\cosh\left(  \beta
Jq_{j}^{1/2}x\right)  ,
\end{equation}
and $1/\beta=k_{B}T$, where $k_{B}$ is the Boltzmann constant and $T$ is the
absolute temperature.

We then analyzed this two-dimensional mapping, given by equations (\ref{eq3})
and (\ref{eq4}), both analytically and by numerical methods. It is interesting
to have in mind the $T-K$ phase diagram obtained by da Costa and collaborators
for the corresponding BEGsg model \cite{daCosta97}. We find two kinds of fixed
points of the recursion relations. The trivial fixed point, given by $q^{\ast
}=0$, $p^{\ast}\neq0$, corresponds to the paramagnetic \textbf{P} phase. A
second, and non trivial, fixed point, given by $q^{\ast}\neq0$, $p^{\ast}%
\neq0$, corresponds to the spin-glass phase \textbf{SG}. As long as $K$ is
positive, there are only these two kinds of fixed points. However, for
negative values of $K$, the repulsive character of this parameter plays a
dominant role at low temperatures. Besides the \textbf{P} and \textbf{SG}
fixed points, we also find two $2$-cycles, which correspond to phases already
found by da Costa and collaborators \cite{daCosta97}: (i) At sufficiently high
temperatures, there appears an antiquadrupolar \textbf{AQ} solution, given by
the two-cycle $p_{1}^{\ast}\neq p_{2}^{\ast}$ and $q_{1}^{\ast}=q_{2}^{\ast
}=0$; (ii) As the temperature decreases, there is an antiquadrupolar
spin-glass solution \textbf{AQG}, corresponding to the two-cycle $(q_{1}%
^{\ast},p_{1}^{\ast})\neq(q_{2}^{\ast},p_{2}^{\ast})\neq(0,0)$.

The standard linear analysis of stability of these single fixed points and
two-cycles leads essentially to the same $T-K$ phase diagram as previously
found by da Costa and coauthors \cite{daCosta97}. The regions associated with
\textbf{P} and \textbf{SG} fixed points are separated by a critical line,%
\begin{equation}
\frac{K}{J}=-\frac{J}{2k_{B}T}-\ln\left(  \frac{2J}{k_{B}T}-2\right)  .
\end{equation}
The regions corresponding to the fixed point \textbf{P} and the \textbf{AQ}
$2$-cycle are separated by the critical border%
\begin{equation}
\frac{K}{J}=-\frac{a k_{B}T}{J}-\frac{J}{2k_{B}T},
\end{equation}
with $a=4.622...$, in agreement with da Costa and coworkers
\cite{daCosta97}.

We also find a critical boundary between the regions of stability of the
\textbf{AQ} and \textbf{AQG} $2$-cycles, as well as between the regions of
stability of the \textbf{SG} fixed point and the \textbf{AQG} $2$-cycle. These
critical boundaries are identical to the respective critical lines for the
\textbf{AQ -- AQG} and \textbf{SG -- AQG} phase transitions, as obtained for
the long-range spin-glass model \cite{daCosta97}.

At this point, our findings can be summarized as follows. We have found
\textbf{P}, \textbf{SG} and \textbf{AQ} solutions, which reproduce the known
results for the corresponding infinite-range model at the replica-symmetric
level (see figure 2 in Ref. \cite{daCosta97}). However, in the $T-K$ phase
diagram, the region in which one anticipates a single \textbf{AQG} structure
is much more complicated. Instead of a stable 2-cycle solution representing
the \textbf{AQG} phase, we find a number of periodic limit cycles (of finite
length) and even chaotic trajectories. The limit cycles can be understood as
modulated structures, which are analogous to results already obtained for an
Ising model on a Cayley tree with competing ferro and antiferromagnetic
interactions between first and second neighbors along the branches of the tree
\cite{Yokoi85}. It should be remarked, however, that neither modulated nor
chaotic phases have been obtained in the replica treatments of the analogous
infinite-range models, even with the introduction of two and three distinct
sublattices. Indeed, in these previous investigations, one finds at most
either $2$-cycle \cite{daCosta97} or $3$-cycle \cite{Katayama00} solutions.

In the present calculation, the occurrence of limit cycles and chaotic
trajectories is more pronounced at low temperatures. In Fig. 1 we show the
order parameters $p$ and $q$ as a function of $K$ for $T=0.07$. In Fig. 2, we
draw the largest Lyapunov exponent at this temperature. The chaotic behavior,
which is associated with a positive Lyapunov exponent, takes place at a small
range of values of $K$ (roughly for $-7.5<K<-6.0$).\bigskip

\begin{figure}[!ht]
\centering
\includegraphics[angle=-90,scale=.33]{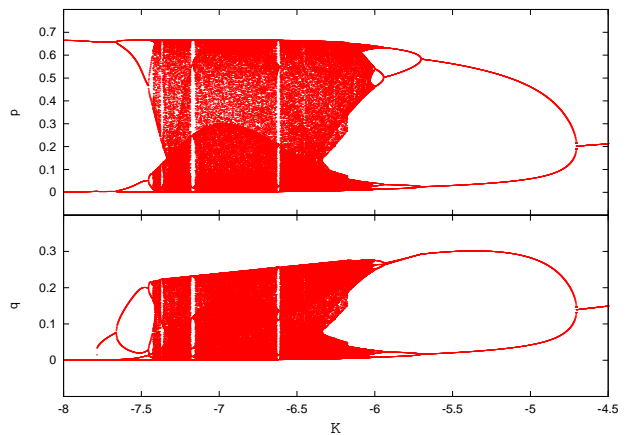}
\caption{(Color online)  Behavior of the order parameters $p$ and $q$ as function of $K$, at
$T=0.07$.}
\label{fig1}
\end{figure}

\begin{figure}[!ht]
\centering
\includegraphics[scale=.6]{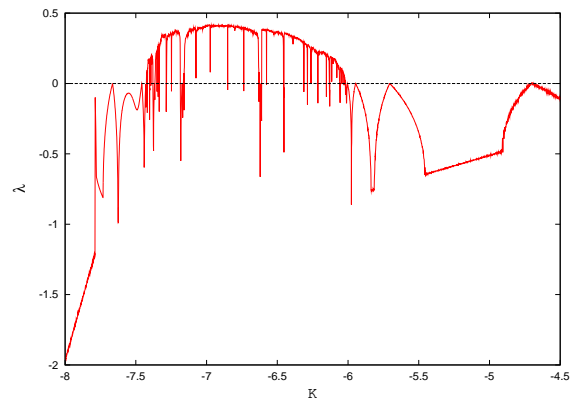}
\caption{(Color online)  Largest Lyapunov exponent as a function of $K$, at $T=0.07$}
\label{fig2}
\end{figure}

In Figs. 3 and 4, we illustrate our numerical findings for $T=0.05$. At this
temperature, as it can be seen from the largest Lyapunov exponent, the chaotic
behavior seems to extend to $K\rightarrow-\infty$. Also, the limit cycles tend
to be much less numerous. According to some preliminary investigations, this
peculiar behavior is not restricted to $T=0.05$, but it does hold at lower
temperatures, down to absolute zero. It is remarkable that such a complex
behavior has been observed for a model on a Cayley tree with only
nearest-neighbor interactions.\bigskip

\begin{figure}[!ht]
\centering
\includegraphics[angle=-90,scale=0.33]{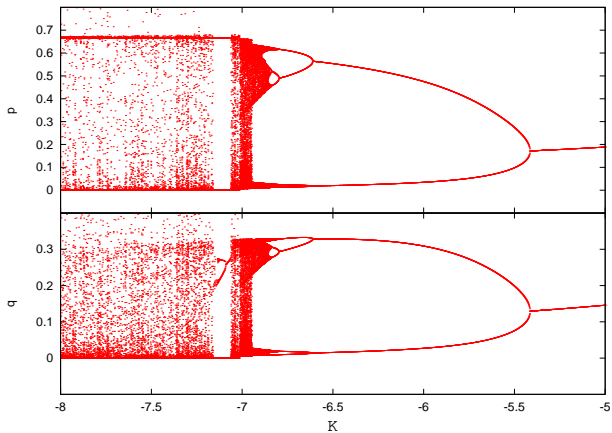}
\caption{(Color online)   Behavior of the order parameters $p$ and $q$ as function of $K$, at
$T=0.05$.}
\label{fig3}
\end{figure}

\begin{figure}[!ht]
\centering
\includegraphics[scale=.6]{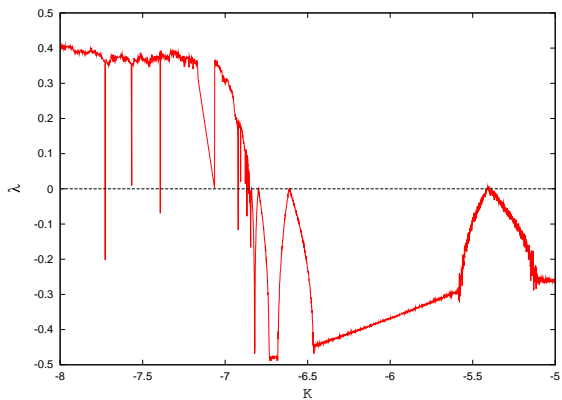}
\caption{(Color online)  Largest Lyapunov exponent as a function of $K$, at $T=0.05$.}
\label{fig4}
\end{figure}

This surprising behavior of the BEGsg on the Cayley tree is an indication that
the more subtle features of the phase diagrams are very difficult to be
obtained by the standard replica solutions of the corresponding infinite-range
models \cite{Crisanti02}\cite{daCosta97}\cite{Katayama00}. These findings can
also provide a useful guide for the interpretation of numerical simulations,
as in some recent Monte Carlo calculations \cite{Paoluzzi10}, and for the
modeling of real systems, as the ferroelastic alloy $Ti_{50}\left(
Pd_{50-x}Cr_{x}\right)  $ in the presence of disorder \cite{Vasseur12}.

In conclusion, the Cayley tree is useful to investigate the spin-glass
properties of a short-range spin-$1$ Ising model. In the ferromagnetic case,
it is well know that the solutions deep inside the tree correspond to the
Bethe approximation. In this note we report a numerical analysis, for an
adequately chosen version of the BEGsg model, and using the recursion
relations obtained by Thompson and collaborators many years ago. The subtle
details of our findings, including long-period cycles and chaotic behavior,
are related to the high degree of frustration of the BEGsg model. All of these
points, as well as some possible connections with more realistic finite-range
spin-glass models at finite temperature, are planned to be further investigated.

F.A. da Costa would like to thank G. M. Viswanathan for helpful comments.

\end{document}